\documentclass{frontiersSCNS} 

\usepackage{url,lineno}
\usepackage{hhline}
\usepackage{multicol}
\usepackage{bm}
\usepackage{dcolumn}
\usepackage{subfigure}
\usepackage{graphicx}
\usepackage{hhline}

\def\keyFont{\fontsize{8}{11}\helveticabold }
\def\firstAuthorLast{Orchard {et~al.}} 
\def\Authors{Garrick Orchard\,$^{1,2,*}$, Ajinkya Jayawant\,$^{3}$, Gregory Cohen\,$^{4}$, and Nitish Thakor\,$^{1}$}
\def\Address{$^{1}$Singapore Institute for Neurotechnology (SINAPSE), National University of Singapore, Singapore\\
$^{2}$Temasek Labs (TLab), National University of Singapore, Singapore\\
$^{3}$Department of Electrical Engineering, Indian Institute of Technology (IIT) Bombay, Mumbai, India\\
$^{4}$MARCS Institute, University of Western Sydney, Penrith NSW, Australia}
\def\corrAuthor{Garrick Orchard}
\def\corrAddress{Singapore Institute for Neurotechnology (SINAPSE), 28 Medical Drive, \#05-COR, 117456, Singapore}
\def\corrEmail{garrickorchard@nus.edu.sg}

\begin{document}
\onecolumn
\firstpage{1}

\title[Spiking datasets for Neuromorphic Vision]{Converting Static Image Datasets to Spiking Neuromorphic Datasets Using Saccades}
\author[\firstAuthorLast ]{\Authors}
\address{}
\correspondance{}
\extraAuth{}
\topic{Benchmarks and Challenges for Neuromorphic Engineering}

\maketitle


\begin{abstract}
Creating datasets for Neuromorphic Vision is a challenging task. A lack of available recordings from Neuromorphic Vision sensors means that data must typically be recorded specifically for dataset creation rather than collecting and labelling existing data. The task is further complicated by a desire to simultaneously provide traditional frame-based recordings to allow for direct comparison with traditional Computer Vision algorithms. Here we propose a method for converting existing Computer Vision static image datasets into Neuromorphic Vision datasets using an actuated pan-tilt camera platform. Moving the sensor rather than the scene or image is a more biologically realistic approach to sensing and eliminates timing artifacts introduced by monitor updates when simulating motion on a computer monitor. We present conversion of two popular image datasets (MNIST and Caltech101) which have played important roles in the development of Computer Vision, and we provide performance metrics on these datasets using spike-based recognition algorithms. This work contributes datasets for future use in the field, as well as results from spike-based algorithms against which future works can compare. Furthermore, by converting datasets already popular in Computer Vision, we enable more direct comparison with frame-based approaches.
\tiny
 \keyFont{\section{Keywords:} Neuromorphic Vision, Computer Vision, Benchmarking, Datasets, Sensory Processing} 
\end{abstract}

\section{Introduction}

Benchmarks, challenges, and datasets have played an important role in the maturation of frame-based Computer Vision \citep{Cheston1}. Quantitative evaluation of algorithms on common datasets and using common metrics allows for a fair and direct comparison between works. This ability to directly compare results encourages competition and motivates researchers by giving them a state-of-the-art target to beat. The importance of datasets extends beyond evaluating and comparing algorithms. Datasets also provide easy access to data for researchers, without which they would be required to gather and label their own data, which is a tedious and time-consuming task.

The task of gathering data is especially tedious for those working in Neuromorphic Vision. A lack of publicly available Neuromorphic data means that Neuromorphic researchers must record their own data, which is in contrast to frame-based Computer Vision, where datasets can be constructed by assembling samples from an abundance of publicly accessible images. Although the barrier to acquiring Neuromorphic Vision sensors has recently been lowered significantly by commercialization of sensors by iniLabs \citep{DVStobi} \footnote{\url{http://www.inilabs.com/}}, a lack of publicly available Neuromorphic Vision data and datasets persists.

The shortage of good datasets for Neuromorphic Vision is well recognised by the community and is in part a catalyst for the Frontiers special topic in which this paper appears. In a separate article in this same special topic we discuss the characteristics of a good dataset, the roles they have played in frame-based Computer Vision, and how lessons learnt in Computer Vision can help guide the development of Neuromorphic Vision. In this paper we focus on creation of Neuromorphic Vision datasets for object recognition.

An important characteristic of a good dataset is that it should be large and difficult enough to cause an algorithm to ``fail" (achieve significantly less than 100\% accuracy). Achieving 100\% accuracy on a dataset sounds impressive, but it does not adequately describe an algorithm's accuracy, it only provides a lower bound. A more accurate algorithm would also achieve 100\% on the same dataset, so a more difficult dataset is required to distinguish between the two algorithms. To ensure the longevity of a dataset, it should be sufficiently difficult to prevent 100\% accuracy from being achieved even in the face of significant algorithmic improvements.

However, many existing Neuromorphic Vision datasets have not been introduced with the aim of providing a long lived dataset. Rather, they have been introduced as a secondary component of a paper describing a new algorithm \citep{hfirst, Perez-Carrasco2013b}. These datasets are introduced only to serve the primary purpose of their paper, which is to show how the algorithm performs, and near 100\% accuracy on the dataset is soon achieved by subsequent improved algorithms.

In this paper our primary aim is to introduce two new Neuromorphic Vision datasets with the goal that they will remain useful to the Neuromorphic community for years to come. Although we provide recognition accuracy of existing algorithms on the datasets, we do so only to provide an initial datapoint for future comparisons. We do not concern ourselves with modifying or improving the algorithms in this paper.

Rather than starting from scratch to record our own datasets, we leverage the existence of well established Computer Vision datasets. By converting Computer Vision datasets to Neuromorphic Vision datasets, we save ourselves considerable time and effort in choosing and collecting subject matter. Furthermore, as we show in Section~\ref{sec:conversion_process}, the conversion process can be automated with a Neuromorphic sensor recording live in-the-loop. Using datasets well known to Computer Vision also ensures easier comparison between communities. The two Computer Vision datasets we have chosen are MNIST \citep{Lecun1998} \footnote{\url{http://yann.lecun.com/exdb/mnist/}} and Caltech101 \citep{Fei-Fei2007} \footnote{\url{http://www.vision.caltech.edu/Image_Datasets/Caltech101/}}. Each of these datasets is intended to play a different role described below. We use the names ``MNIST" and ``Caltech101" to refer to the original Computer Vision datasets, and the names ``N-MNIST" and ``N-Caltech101" to refer to our Neuromorphic versions.

MNIST contains only 10 different classes, the digits 0-9. The examples in the database are small (28$\times$28 pixels), so it can easily be downloaded, copied, and distributed. The small example size also reduces processing time, allowing for rapid testing and iteration of algorithms when prototyping new ideas. An example of the use of MNIST to explore new ideas can be found in Geoffrey Hinton's online presentation on ``Dark Knowledge" \footnote{\url{https://www.youtube.com/watch?v=EK61htlw8hY}}. We intend for N-MNIST to play a similar role in Neuromorphic Vision and have therefore intentionally kept the recorded examples at the same small scale of 28 $\times$ 28 pixels.

Caltech101 is a much more difficult dataset containing 100 different object classes, plus a background class. The images themselves are much larger, averaging 245 pixels in height and 302 pixels in width. While MNIST can be seen as a scratchpad on which to prototype ideas, Caltech101 provides a far more difficult challenge. We acknowledge that Caltech101 is now considered an easy dataset for Computer Vision given the very advanced state of Computer Vision algorithms, but we foresee it posing a significant challenge to the less mature field of Neuromorphic Vision.

Examples of other early Neuromorphic datasets for recognition include the four class card pip dataset from \citep{Perez-Carrasco2013b}, the 36 character dataset from \citep{hfirst}, the four class silhouette orientation dataset from \citep{Perez-Carrasco2013b}, and the 3 class posture dataset from \citep{Zhao2014}. Accuracy on these datasets is already high and they each include only a few stimulus samples (less than 100).

Others have attempted conversion of static images to Neuromorphic data, but the conversion images proves difficult because the fundamental principle underlying Neuromorphic sensors is that they respond only to changes in the scene. Some have approached the problem using simulation. \citep{Masquelier2007} assume spike times to be proportional to local image contrast for a static image, while \citep{OConnor2013a} simulate image motion to create a spike sequence. However, simulations do not realistically approximate the noise present in recordings, which can take the form of spurious events, missing events, and variations in event latency.

Arguably the most complete dataset created thus far is the ``MNIST-DVS" dataset \footnote{\url{http://www2.imse-cnm.csic.es/caviar/MNISTDVS.html}}, which is recorded from an actual sensor \citep{Serrano-Gotarredona2013} viewing MNIST examples moving on a computer monitor. However, this approach is also problematic because motion on a monitor is discontinuous, consisting of discrete jumps in position at each monitor update. These discontinuities are clearly visible in the data as shown later in Fig.~\ref{fig:FourierBernabe}. Furthermore, the MNIST-DVS dataset only converted a 10 000 sample subset of the 70 000 sample in MNIST, preventing Neuromorphic researchers from directly comparing their algorithms to Computer Vision using the same test and training splits. The MNIST-DVS examples have also been upscaled to 3 different scales, resulting in larger examples which are more computationally intensive to process than the smaller recordings we present.

Our approach to converting images uses static images on a computer monitor and instead moves the sensor itself, as described in Section~\ref{sec:conversion_process}. Our approach bears resemblance to retinal movements observed in primate and human experiments \citep{Engbert2006}. These movements are subconscious, they are present even when trying to fixate on a point, and these movements are thought to play an important role in recognition in the primate visual system.

In the rest of this paper, we start off with describing our image conversion process in Section~\ref{sec:conversion_process} and using it to convert the MNIST and Caltech101 datasets. In Section~\ref{sec:Properties} we show examples of recordings and describe some of the properties of the recorded datasets. In Section~\ref{sec:Recognition} we briefly present recognition accuracies on the datasets using previously published algorithms before wrapping up with discussion in Section~\ref{sec:Discussion}.

\section{Converting Static Images to Neuromorphic Recordings}
\label{sec:conversion_process}

In this section we describe the principle behind our image conversion technique (Section~\ref{sec:fourier_desc}), the hardware and software design of a system to implement this technique (Section~\ref{sec:sub:design}), the specific parameters used by the system for conversion of MNIST and Caltech101 (Section~\ref{sec:sub:params}), and information on how to obtain and use the resulting datasets (Section~\ref{sec:sub:formats}).

\subsection{Approach to Static Image Conversion}
\label{sec:fourier_desc}

As discussed in the previous section, creating Neuromorphic databases from existing frame based datasets saves us time in collecting subject matter and creates a dataset familiar to frame-based Computer Vision researchers, allowing for more direct comparisons between fields. However, the question of how to perform the conversion remains. Below we discuss several possible approaches to performing the conversion and provide the reasoning which led us to our final conversion process.

Neuromorphic Vision sensors are specifically designed such that each pixel responds only to changes in pixel intensity \citep{Posch2014}. These changes can arise either from changes in lighting in the real-world scene, or from the combination of image motion and image spatial gradients. Although one can imagine schemes in which the scene illumination is modified to elicit pixel responses (e.g. turning on the lights), such a process is unnatural in the real world and infeasible in brightly lit outdoor conditions where we would expect performance to be best. We therefore chose to instead use image motion as the mechanism by which to elicit changes in pixel brightness.

Even for a scene of constant brightness, the brightness observed by an individual pixel changes over time as sensor or object motion causes the same pixel to view different parts of the scene. The canonical optical flow constraint describing the change in brightness of an individual point on the image plane can be derived from the image constancy constraint as:
\begin{equation}\label{eq:FlowConstraint}
\begin{array}{l l l}
I_t &= &-I_xV_x - I_yV_y\\
\end{array}
\end{equation}
where $I_t$, $I_x$, and $I_y$ are shorthand for the derivatives of image intensity ($I$) with respect to time ($t$), and $x$ and $y$ spatial co-ordinates on the image plane respectively. $V_x$ and $V_y$ are velocities on the image plane in the $x$ and $y$ directions. The equation describes how changes in pixel brightness ($I_t$) arise as a combination of image motion ($V_x$ and $V_y$) and image spatial gradients ($I_x$ and $I_y$). The image motion in \eqref{eq:FlowConstraint} above is due to relative motion between the sensor and subject matter. The image motion can be described as:
\begin{equation}\label{eq:VisualMotion}
\begin{array}{l l l}
V_x &= &\frac{T_zx-T_x}{z} - \omega_y + \omega_zy + \omega_xxy - \omega_yx^2\\
V_y &= &\frac{T_zy-T_y}{z} + \omega_x - \omega_zx - \omega_yxy + \omega_xy^2\\
\end{array}
\end{equation}
where $T_x$, $T_y$, and $T_z$ are translational velocities of sensor relative to the scene, $\omega_x$, $\omega_y$, and $\omega_z$ are rotational velocities around the sensor axes, and $x$, $y$, and $z$ are co-ordinates of points in the scene relative to the sensor.

Image motion resulting from relative translation between the sensor and subject matter ($T_x$, $T_y$, $T_z$) is dependent on scene depth ($z$), but for a sensor viewing a static 2D image, all points in the image will effectively have the same depth (assuming the 2D image is parallel to the sensor image plane). Therefore, the response ($I_t$) when translating a sensor while viewing a static 2D image differs from the response when translating the sensor while viewing the original 3D scene from which the image was captured. On the other hand, image motion induced by rotation of the camera about its origin ($\omega_x$, $\omega_y$, $\omega_z$) is not dependent on scene depth. For this reason, we decided that the relative motion between the image and camera should take the form of pure rotation about the camera origin.

We chose to record with a real sensor in the loop viewing images on a monitor rather than using pure simulation. Using actual sensor recordings lends the dataset more credibility by inherently including the sensor noise which can be expected in real-world scenarios. We chose to physically rotate the sensor itself rather than rotating the image about the camera origin because it is both more practical and more realistic to a real-world scenario. One could imagine a scenario in which the image rotation around the sensor is simulated on a PC monitor, but motion on a monitor is discontinuous and clearly shows up in recordings.

\begin{figure}
\centering
\includegraphics[width=0.5\columnwidth]{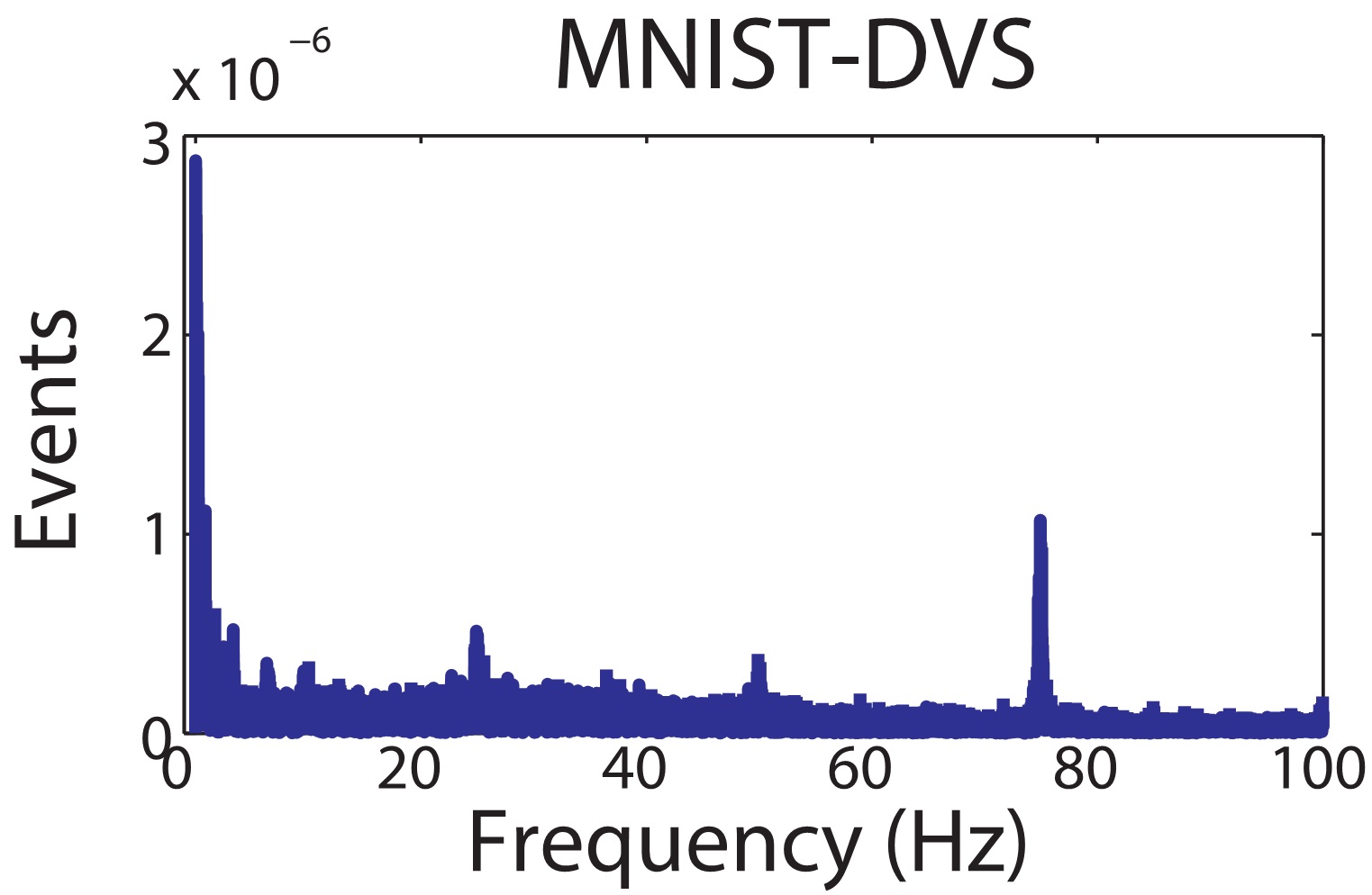}
\caption{A Fourier analysis of the MNIST-DVS dataset showing the temporal frequencies at which events occur. The 0Hz component has been removed, and the energy in the signal has been normalized to 1 $event^2 ms$. Clear peaks are observed at low frequencies due to the slowly varying motion of the digits on the monitor. A significant peak is observed at 75Hz due to the discontinuous motion presented to the sensor as a result of the 75Hz monitor refresh rate.}
\label{fig:FourierBernabe}
\end{figure}

Fig~\ref{fig:FourierBernabe} shows how simulating motion on a monitor affects recordings. The figure shows the Discrete Fast Fourier Transform (DFFT) of data from the MNIST-DVS dataset. For a more accurate DFFT, recordings were randomly selected and concatenated until a recording of length $2^{27}$ microseconds was achieved. A zero vector of $2^{27}$ timesteps was created and any timestep where an event occurred was incremented. The mean value of the vector was then subtracted to remove the DC (0 Hz) component, and the energy in the vector was normalized to 1 $event^2ms$ by dividing by the $l_2$ norm.

Large low frequency components ($<$ 5Hz) can be seen due to the slowly varying motion of the characters on the screen. A significant peak is observed at 75Hz due to discontinuities in the motion caused by the monitor refresh rate (75Hz). Secondary peaks can also be observed at 50Hz and 25Hz.

\subsection{Design of Static Image Conversion System}
\label{sec:sub:design}
\subsubsection{Hardware Design}

\begin{figure}
\centering
\begin{tabular}{c|c}
\subfigure[]{\includegraphics[width=0.15\columnwidth]{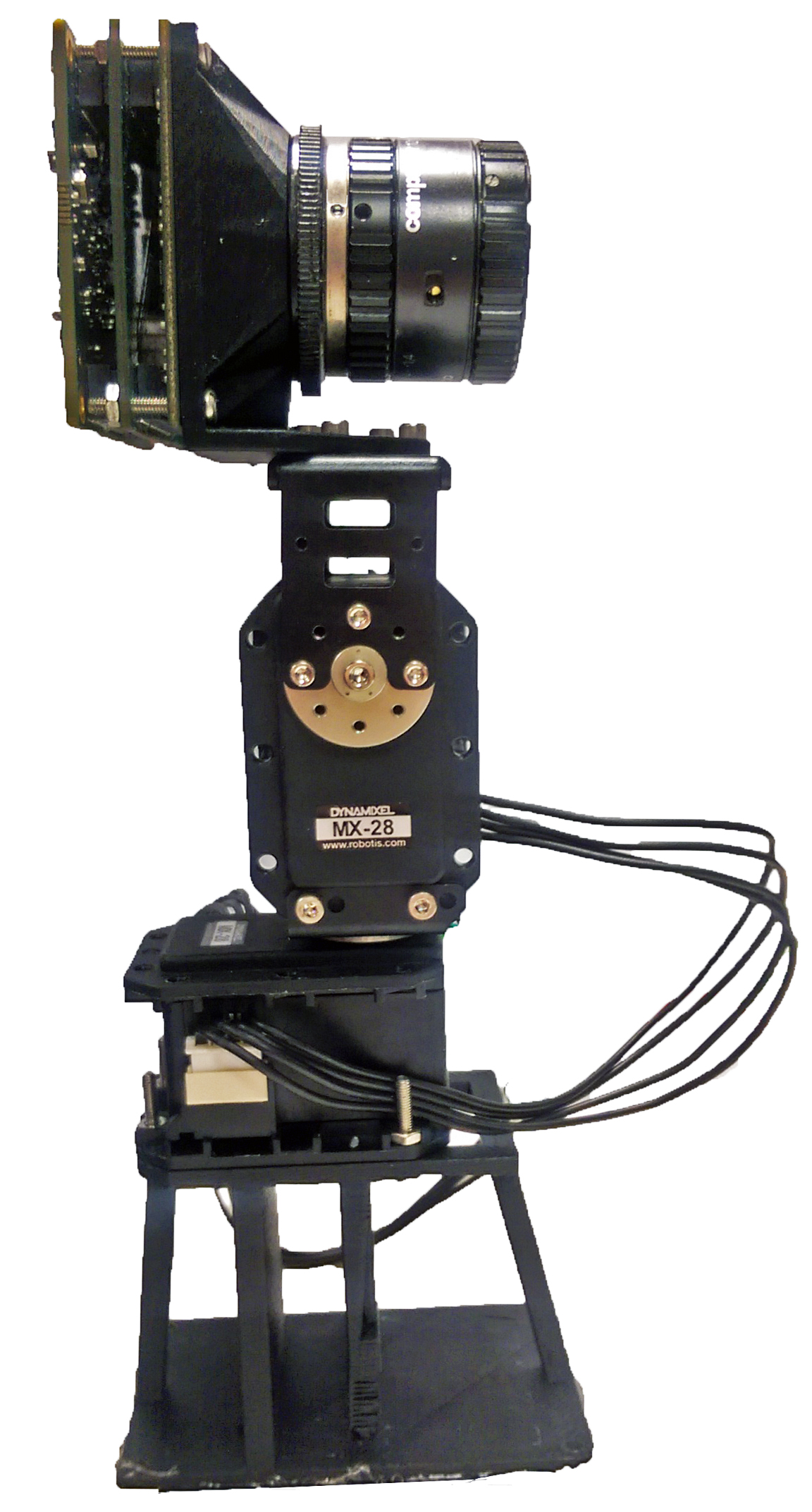}}
&
\subfigure[]{\includegraphics[width=0.8\columnwidth]{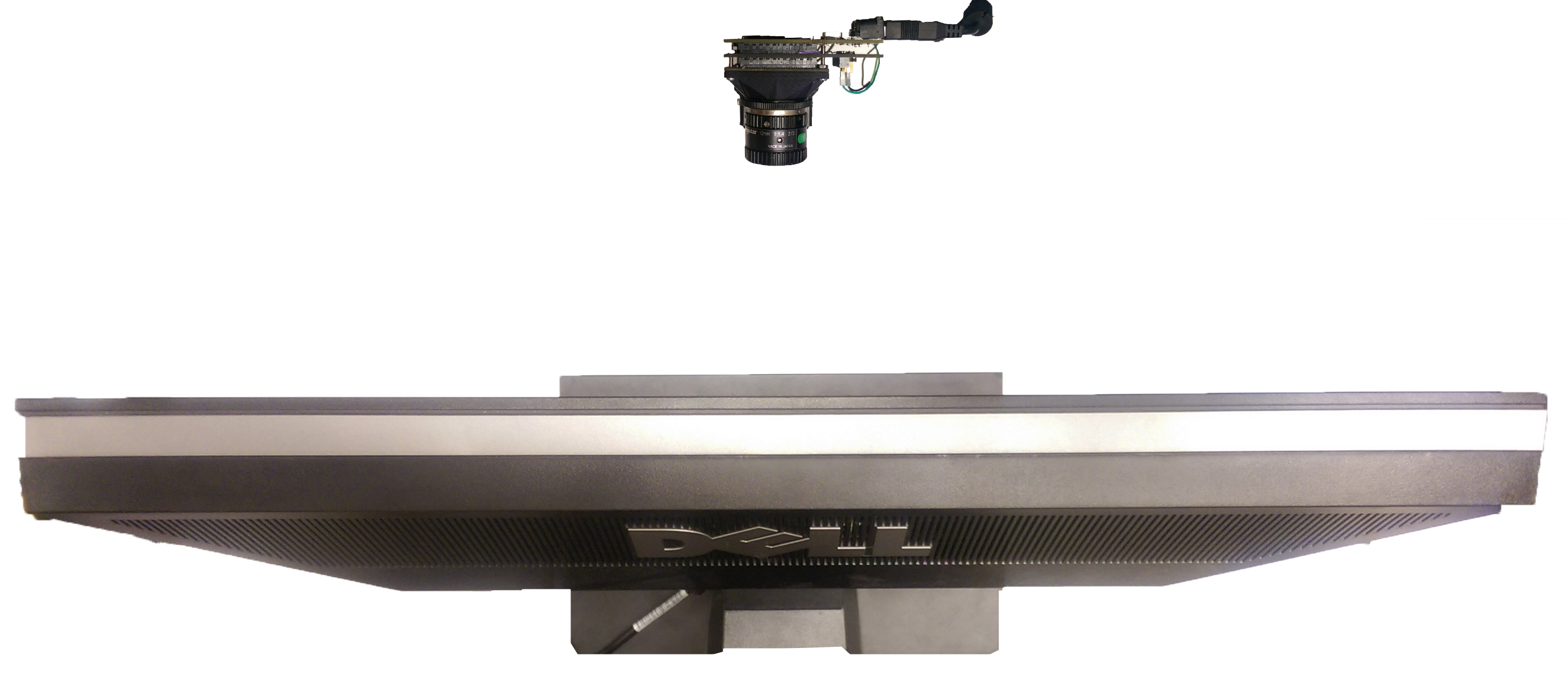}}
\end{tabular}
\caption{(a) A picture of the ATIS mounted on the pan tilt unit used in the conversion system. (b) The ATIS placed viewing the LCD monitor.}
\label{fig:Hardware}
\end{figure}

Our conversion system relies on the Asynchronous Time-based Image Sensor (ATIS) \citep{ATIS} for recording. To control motion of the ATIS, we constructed our own pan-tilt mechanism as shown in Fig.~\ref{fig:Hardware}. The mechanism consists of two Dynamixel MX-28 motors \footnote{\url{http://www.trossenrobotics.com/dynamixel-mx-28-robot-actuator.aspx}} connected using a bracket. Each motor allows programming of a target position, speed, and acceleration. A custom housing for the ATIS including lens mount and a connection to the pan-tilt mechanism was 3D printed. The motors themselves sit on a 3D printed platform which gives the middle of the sensor a height of 19cm, high enough to line up with the vertical center of the monitor when the monitor is adjusted to its lowest possible position. The motors interface directly to an Opal Kelly XEM6010 board containing a Xilinx Spartan 6 lx150 Field Programmable Gate Array (FPGA) \footnote{\url{https://www.opalkelly.com/products/xem6010/}} using a differential pair. The Opal Kelly board also serves as an interface between the ATIS and host PC. Whenever a motor command is executed, the FPGA inserts a marker into the event stream from the ATIS to indicate the time at which the motor command was executed. The entire sensor setup was placed at a distance of 23cm from the monitor and enclosed in a cupboard to attenuate the effects of changing ambient light. A Computar M1214-MP2 2/3" 12mm f/1.4 lens \footnote{\url{http://computar.com/product/553/M1214-MP2}} was used.

\subsubsection{Software Design}

\begin{figure}
\centering
\includegraphics[width=0.9\columnwidth]{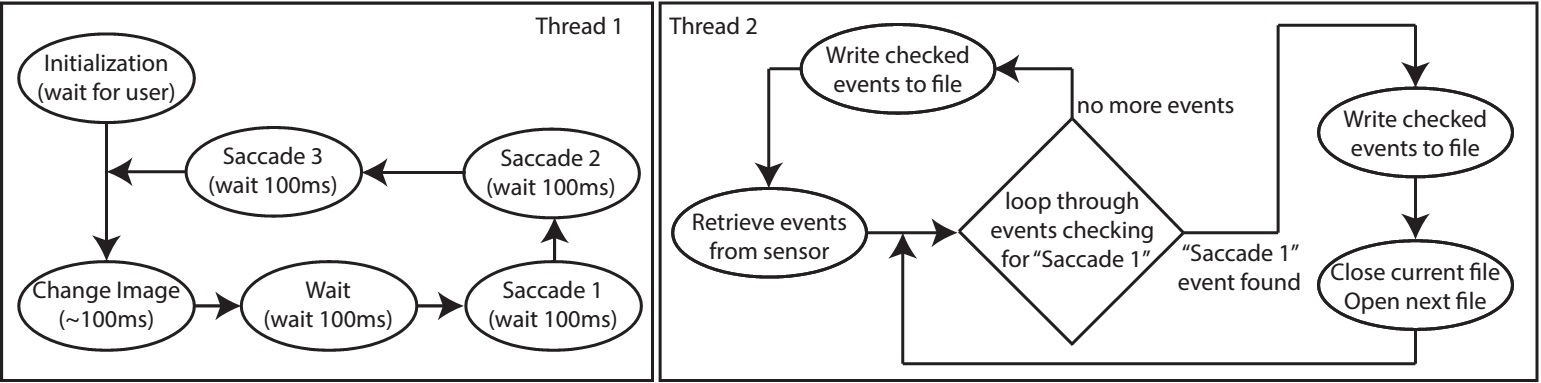}
\caption{Software flow diagram of the PC code showing two threads. Thread 1 (left) controls the timing of microsaccades and changing of images on the screen, while thread 2 (right) controls acquisition of the event data and writing it to disk.}
\label{fig:Software}
\end{figure}

A C\# GUI on the host-PC interfaces with the Opal Kelly board to control the motors and the ATIS. This same C\# GUI also controls the display of images on the monitor and handles recording of data. The GUI has two main threads. The first thread consists of a state machine with 5 different states as shown in Fig.~\ref{fig:Software}.

At the beginning of the \textit{Initialization} state, the directory containing the images to be converted, and the directory to which the output should be written are specified. The GUI parses the image directory and subdirectory for images, and creates an identical directory structure at the output. Then the user uses the grayscale function of the ATIS as visual feedback to modify the scale and position at which images on the monitor will be displayed to ensure that they are centered in the ATIS' field of view and match the desired output size ($pixels^2$) from the ATIS (indicated using an overlay on the ATIS display). Once the user acknowledges that this initialization procedure is complete, the GUI enters the \textit{Change Image} state and the rest of the process is automated.

During the \textit{Change Image} state, the next image to be converted is pushed to the display. A software check is used to enure that the monitor has updated before proceeding. This check prevents a rare occurrence (less than 1 in 50k recordings) in which the monitor update would be significantly delayed. Once the check has passed (\~100ms) the \textit{Wait} state is entered during which a 100ms interrupt timer is initialized and used to transition between subsequent states. During the \textit{Wait} state, the 100ms delay allows time for the sensor to settle after detecting the visual changes incurred by changing the image.

During the \textit{Saccade 1}, \textit{Saccade 2}, and \textit{Saccade 3} states, the commands to execute the 1st, 2nd, and 3rd micro-saccades respectively are sent to the Opal Kelly. After the \textit{Saccade 3} state, the timer interrupt is disabled and the code returns to the \textit{Change Image} state. This process repeats until all images are processed.

A second thread operates in parallel to the first. It pulls ATIS data off the Opal Kelly over USB2.0 and writes it to the corresponding file in the output directories. The thread parses the event stream looking for the marker indicating that \textit{Saccade 1} is about to be executed to determine when to end recording for the current image and begin recording for the next image.

Using this automated process, on average each image takes under 500ms to convert, so the entire MNIST database of 70k images can be converted in under 9.5hrs, and the 8709 image Caltech101 database takes roughly 75 minutes. 
Video of the conversion system in action \footnote{Video of the system recording \url{https://youtu.be/2RBKNhxHvdw}} and videos showing converted N-MNIST \footnote{Video showing N-MNIST data \url{https://youtu.be/6qK97qM5aB4}} and N-Caltech101 \footnote{Video showing N-Caltech101 data \url{https://youtu.be/dxit9Ce5f_E}} examples can be found online.

\subsection{Recording Parameters}
\label{sec:sub:params}
\begin{table}
\textbf{\refstepcounter{table}\label{Table:saccade_parameters} Table \arabic{table}.}{ Motion parameters for the ATIS during each state during the acquisition process}
\begin{center}
\begin{tabular}{|c||D..{3}| *{3}{|D..{3.1}|D..{3.1}|}|}
  \hline
    \rule{0pt}{2ex}   &\multicolumn{1}{c||}{\textbf{Start time}} &\multicolumn{2}{c||}{\textbf{Start (deg)}}  & \multicolumn{2}{c||}{\textbf{End (deg)}}   & \multicolumn{2}{c|}{\textbf{Speed (deg/s)}} \\ \hline
            \rule{0pt}{2ex} \textbf{State} & \multicolumn{1}{c||}{\textbf{(ms)}} & \textbf{x}    &  \textbf{y}                        &   \textbf{x}           &     \textbf{y}     &   \textbf{x}       &   \textbf{y}           \\ \hline
   \rule{0pt}{2ex}\textbf{Change Image} & 0     & -0.5 &  0.5                      &   -0.5 &  0.5      &   0      &   0          \\ \hline
   \rule{0pt}{2ex}\textbf{Wait}         & 100   & -0.5 &  0.5                      &   -0.5 &  0.5      &   0      &   0          \\ \hline
   \rule{0pt}{2ex}\textbf{Saccade 1}    & 200   & -0.5 &  0.5                      &   0           & -0.5      &   10      &   20          \\ \hline
   \rule{0pt}{2ex}\textbf{Saccade 2}    & 300   & 0    &  -0.5                     &   0.5         & 0.5       &   10      &   20          \\ \hline
   \rule{0pt}{2ex}\textbf{Saccade 3}    & 400   & 0.5  &  0.5                      &   -0.5        & 0.5       &   20      &   0           \\
  \hline
\end{tabular}
\end{center}
\end{table}

To ensure consistent stimulus presentation, the same sequence of 3 micro-saccades tracing out an isosceles triangle was used on each image. This pattern ensures that the sensor finishes in the correct starting position for the next image. It also ensures that there is motion in more than 1 direction which is important for detecting gradients of different orientations in the image. Saccading back and forth between only two points would produce very weak responses to gradients in a direction perpendicular to the line joining those points. Micro-saccade onset times are spaced 100ms apart and the parameters used for each micro-saccade are shown in Table~\ref{Table:saccade_parameters}. Analog bias parameters used for the ATIS chip during recording are available online with the dataset downloads.

To create the N-MNIST dataset, MNIST images were resized to ensure that each image projects to 28$\times$28 pixels on the ATIS (since 28$\times$28 pixels is the size of the original MNIST images). Original images in the Caltech101 dataset vary in both size and aspect ratio. The approach used in \citep{Serre2007a} was adopted for resizing Caltech101 images before recording. Each image was resized to be as large as possible while maintaining the original aspect ratio and ensuring that width (x-direction) does not exceed 240 pixels and height (y-direction) does not exceed 180 pixels.

\subsection{File Formats}
\label{sec:sub:formats}

The full datasets, as well as code for using them, can be accessed online \footnote{\url{http://www.garrickorchard.com/datasets}}. A separate directory exists for each class, and for N-MNIST, separate testing and training directories exist. Each example is saved as a separate binary file with the same filename as the original image.

Each binary file contains a list of events, with each event occupying 40 bits. The bits within each event are organized as shown below. All numbers are unsigned integers.
\begin{itemize}
\item bit 39 - 32: Xaddress (in pixels)
\item bit 31 - 24: Yaddress (in pixels)
\item bit 23: Polarity (0 for OFF, 1 for ON)
\item bit 22 - 0: Timestamp (in microseconds)
\end{itemize}

The Caltech101 dataset also comes with two types of annotations. The first is bounding boxes containing each object. The second is a contour outline of the object. With the online dataset, we supply both of these types of annotations, derived from the original Caltech101 annotations.

\section{Dataset Properties}
\label{sec:Properties}

Table~\ref{Table:stats} shows some basic properties of each dataset. As expected, the larger Caltech101 images generate more events than the MNIST images, but for both datasets there is a roughly equal ratio of ON events to OFF events. The mean value of event x-addresses and y-addresses depends on both the image content and the image size, and can therefore be used for classification (described in Section~\ref{sec:Recognition}). The range of event x-addresses and y-addresses depends only on the size of the original input images, which is the same for all MNIST images, but varies for Caltech101 images.

\begin{table}
\textbf{\refstepcounter{table}\label{Table:stats} Table \arabic{table}.}{ Statistics of the dataset recordings which were used for classification}
	\begin{center}
		\begin{tabular}{|l||c|c||c|c|}  \hline
			\rule{0pt}{2ex}\textbf{Dataset} &\multicolumn{2}{c|}{\textbf{N-MNIST}} & \multicolumn{2}{c|}{\textbf{N-Caltech101}}  \\ \hline
            \rule{0pt}{2ex}& \multicolumn{2}{c||}{\textbf{Object Categories}}& \multicolumn{2}{c|}{\textbf{Training Set}}\\ \hline
			\rule{0pt}{2ex}\textbf{Statistic}  &\textbf{Mean} & \textbf{$\sigma$} & \textbf{Mean} & \textbf{$\sigma$}\\ \hline
			\rule{0pt}{2ex}\textbf{ON Events}  & 2084 & 574 & $56936$ & $28039$     \\ \hline
			\rule{0pt}{2ex}\textbf{OFF Events}& 2088 & 623  & $58180$ & $30021$     \\ \hline
            \rule{0pt}{2ex}\textbf{X Mean}   & 17.66 & 5.05 & $100.72$ & $57.78$        \\ \hline
			\rule{0pt}{2ex}\textbf{Y Mean}   & 18.10 & 6.38 & $81.23$  & $46.15$  \\ \hline			
            \rule{0pt}{2ex}\textbf{X Range}  & 34.00 & 0.00 & $198.53$ & $43.08$  \\ \hline
			\rule{0pt}{2ex}\textbf{Y Range}  & 34.00 & 0.00 & $155.88$ & $26.76$ \\ \hline
		\end{tabular}
	\end{center}
\end{table}

\begin{figure}
\centering
\includegraphics[width=\columnwidth]{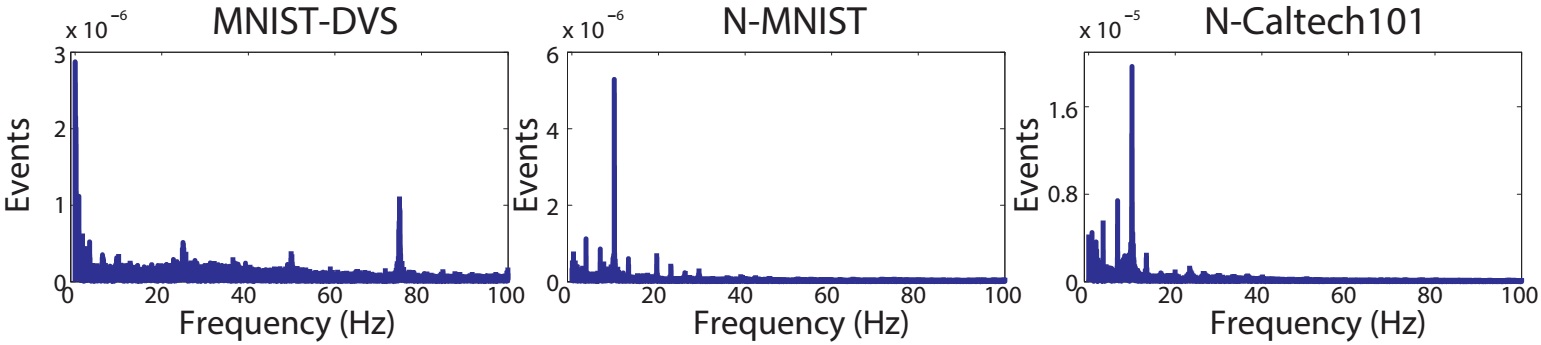}
\caption{A Fourier analysis showing the frequency at which events are elicited during recording for each dataset created using the method described in Section~\ref{sec:fourier_desc}. The leftmost figure is a repeat of Fig.~\ref{fig:FourierBernabe} showing MNIST-DVS with a peak at 75Hz. The middle and right show the N-MNIST and N-Caltech101 datasets respectively. The two rightmost examples show no peak at 75Hz. They have a strong 10Hz peak due to the 10Hz frequency of saccades (100ms each), and a strong peak at 3.3Hz due to the length of each recording (300ms). Harmonics of 3.3Hz and 10Hz can also be seen. A similar 150Hz harmonic exists in the MNIST-DVS data but is not shown in order to improve visibility for lower frequencies. 
}
\label{fig:Fourier_N}
\end{figure}

Fig.~\ref{fig:Fourier_N} presents a Fourier analysis showing the temporal frequency of events using the method described in Section~\ref{sec:fourier_desc}. In all three recordings, the strongest frequencies correspond to frequencies of visual motion observed by the sensor. Making images move on a computer monitor (left) results in an intended strong low frequency component due to the slowly changing motion on the screen, but a second, unintended strong frequency component is present at the monitor refresh rate (75Hz) due to the discontinuous nature of the motion. For the N-MNIST and N-Caltech101 datasets where the sensor is moved instead, strong components are observed at frequencies corresponding to the motor motion (10Hz), frequencies corresponding to the length of recordings (3.3Hz), and harmonics of these frequencies. Harmonics of 75Hz are present in the MNIST-DVS frequency spectrum, but are not shown.

\begin{figure}
\centering
\includegraphics[width=0.9\columnwidth]{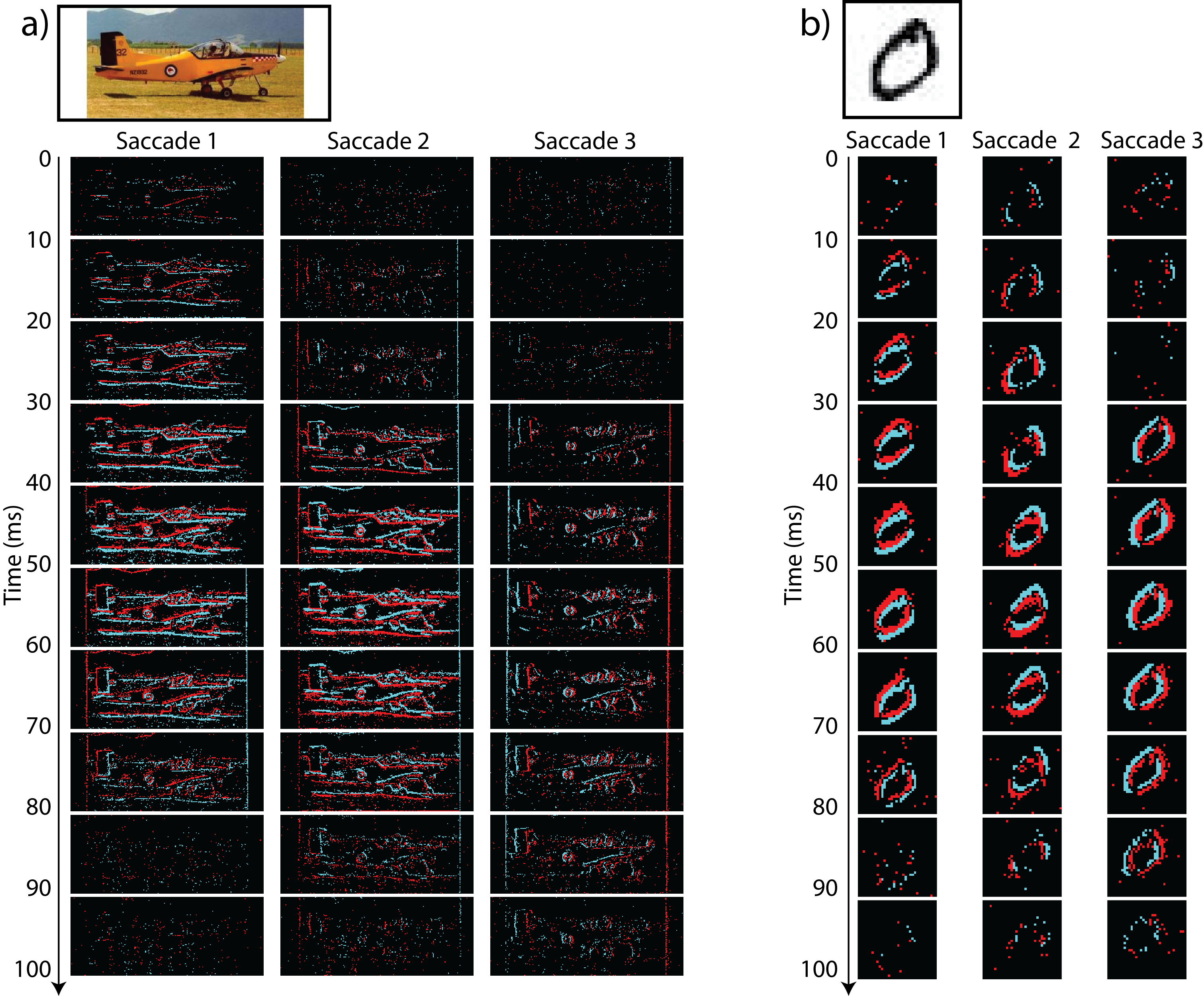}
\caption{Typical recordings for Caltech101 (a) and MNIST (b). The original images are shown at the top, and the recorded events obtained during each saccade are shown below. Each event image shows 10ms worth of data. Black regions indicate no events, red indicates ON events, and blue indicates OFF events. The third saccade is the shortest in distance and therefore generates the fewest events. Fewer events are recorded near the start and end of each saccade when the ATIS is moving slowest.}
\label{fig:Recordings}
\end{figure}

Fig.~\ref{fig:Recordings} shows one example recording from each of the N-Caltech101 (left) and N-MNIST (right) datasets. The original images are shown at the top, with neuromorphic recordings shown below. Each of the neuromorphic subimages contains 10ms of events. In each case the most events are present near the middle of a saccade when the sensor is moving fastest.

The airplane image highlights a few properties of the dataset. For Caltech101 some images have unusual aspect ratios. This is especially obvious for the airplane images which are very wide, with many including additional white space at the sides of the images (as is the case with this example). The border of the image will generate events during the saccade, but care has been taken to remove the image borders and any events occurring outside the borders from the dataset. However, borders contained within the image (such as in this example) have intentionally been left in place.

The airplane example is dominated by strong vertical gradients ($I_y$, horizontal lines) and therefore generates far fewer events in response to \textit{Saccade 3} which is a pure rotation ($\omega_y$), about the y-axis. The smaller number of events is predicted by \eqref{eq:VisualMotion} which indicates that y-axis rotation results in large $V_x$ but small $V_y$ visual flow. \eqref{eq:FlowConstraint} shows that a low value of $V_y$ will attenuate the effect of the strong vertical gradients $I_y$ on $I_t$, and therefore result in fewer output events.

\begin{figure}
\centering
\includegraphics[width=\columnwidth]{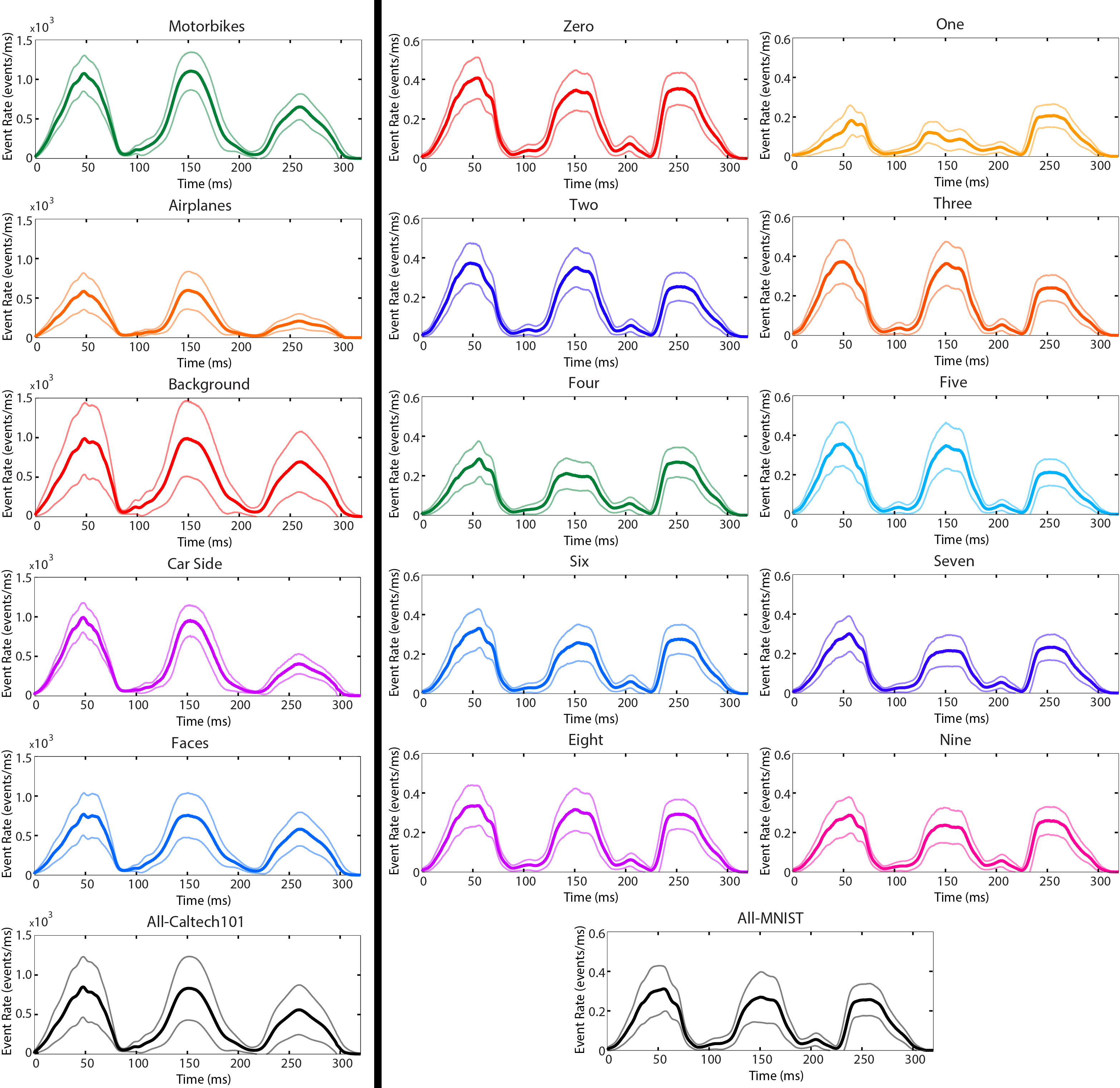}
\caption{The mean (solid) and standard deviation (transparent) event rates per $\mu s$ for popular N-Caltech101 categories (left) and the N-MNIST dataset (right). The three peaks in each plot correspond to the three saccades. As expected from \eqref{eq:VisualMotion}, the maximum event rates occur near the middle of each saccade when the rotational velocity is highest.}
\label{fig:rates}
\end{figure}

Fig.~\ref{fig:rates} shows the average event rate (in events per millisecond) across time for popular classes in Caltech101. The \textit{Faces} and \textit{Background} categories both show slightly lower event rates during the third saccade because the third saccade is the shortest (in angle) and slowest. The \textit{Car Side}, \textit{Airplanes}, and \textit{Motorbikes} categories all show significantly lower event rates during the third saccade due to strong vertical gradients in the images. The \textit{Airplanes} category shows a significantly lower event rate throughout the recording due to the unusual short-and-wide aspect ratio of the images which results in a smaller overall image area when scaled to fit within the 240$\times$180 pixel viewing area as described in Section~\ref{sec:conversion_process}.

For the N-MNIST recordings, the digit ``1" has a significantly higher event rate during the third saccade due to the presence of strong horizontal gradients ($I_x$) and absence of strong vertical gradients ($I_y$) in the images.

\section{Recognition}
\label{sec:Recognition}

Here we briefly present recognition results using existing algorithms to provide an initial recognition accuracy target to beat. We apply these algorithms ``as is" without any modification because development and tuning of recognition algorithms is beyond the scope of this paper. In each case, we refer the reader to the original algorithm papers for detailed description of the algorithm. Three approaches to recognition were used. The first uses statistics of the recordings (such as the number of events in an example), the second uses the Synaptic Kernel Inverse Method (SKIM) \citep{Tapson2013}, and the third uses the HFIRST algorithm \citep{hfirst}. Each of these approaches is described in a subsection below.

\subsection{Recognition by statistics}
\label{sec:sub:kNN}

For each recording, eleven different statistics were calculated. These are statistics are:
\begin{enumerate}
\item The total number of events
\item The number of ON events
\item The number of OFF events
\item The ratio of ON to OFF events
\item The mean X address of events
\item The mean Y address of events
\item The standard deviation in X address of events
\item The standard deviation in Y address of events
\item The maximum X address of events
\item The maximum Y address of events
\end{enumerate}

For classification, the above statistics are treated as features and a k-Nearest Neighbour (kNN) classifier with $k = 10$ is used to determine the output class. For N-MNIST we test using the entire test set, in each case finding the 10 nearest neighbours in the training set.

For N-Caltech101, the number of samples in each class ranges from 31 (inline skate) to 800 (airplanes). To ensure the same number of test and training samples were used for each class, we always used 15 training samples and 15 test samples per class.

\subsection{Synaptic Kernel Inverse Method (SKIM)}
\label{sec:sub:SKIM}

The SKIM was used to form a classifier for both the N-MNIST and N-Caltech datasets, making use of the standard network configuration presented in the original SKIM paper \citep{Tapson2013}. A 1ms timestep was used throughout and each pixel is treated as an individual input channel. Alpha functions with delays were used as the post-synaptic potentials in the hidden layer, with a sigmoidal non-linearity at the output of each hidden layer node. The maximum values for the delays and durations of the alpha functions were configured to lie within the time duration of the longest recording (316 ms). Training output patterns consisted of a square pulse of 10ms in length to indicate when the output spike should occur.
All output neurons were trained together, and the neuron achieving the maximum value during the output period was selected as the classifier output.

For the N-MNIST dataset, 2000 hidden layer neurons were used. Training used 10 000 randomly selected samples from the training set, and testing was performed using the full testing set. For the N-Caltech101 dataset, a similar SKIM network was implemented using 5000 hidden layer neurons.

\subsection{HFIRST}

\label{sec:sub:HFIRST}
\begin{table}
  \textbf{\refstepcounter{table}\label{Table:NetworkParameters} Table \arabic{table}.}{ HFIRST parameters used for N-MNIST}
\begin{center}
\begin{tabular}{|c||c|c|c|c|c|}
\hline
\rule{0pt}{2ex} \textbf{Layer}              & ${V_{thresh}}$ & ${{I_l}/{C_m}}$          &${t_{refr}}$     & \textbf{Kernel Size}   & \textbf{Layer Size}  \\ \hline
\rule{0pt}{2ex}\textbf{S1}                  &150            &25                     &5              &7$\times$7$\times$1        & 34$\times$34$\times$12     \\ \hline
\rule{0pt}{2ex}\textbf{C1}                  &1              &0                      &5              &4$\times$4$\times$1        & 9$\times$9$\times$12  \\ \hline
\rule{0pt}{2ex}\textbf{S2}                  &150            &1                      &5              &9$\times$9$\times$12       & 1$\times$1$\times 10$    \\ \hline
\rule{0pt}{2ex}\textbf{C2}                  &1              &0                      &5              &1$\times$1$\times 1$       & 1$\times$1$\times 10$\\ \hline
\rule{0pt}{2ex}\textbf{unit}               &\textbf{mV}    &\textbf{mV/ms}          &\textbf{ms}    &\textbf{synapses}          & \textbf{neurons}     \\ \hline
\end{tabular}
\end{center}
\end{table}

The HFIRST algorithm as described in \citep{hfirst} was only applied to the N-MNIST dataset because application to N-Caltech101 would require extension of the algorithm to handle such large images. The parameters used are shown in Table~\ref{Table:NetworkParameters}. Ten S2 layer neurons were trained, one for each output class. The input synaptic weights for each S2 layer neuron are determined by summing the C1 output spikes from all training samples of the same class. As in the original HFIRST paper, two different classifiers were used. The first is a hard classifier which chooses only the class which generated the most output spikes. The second is a soft classifier which assigns a percentage probability to each class equal to the percentage of output spikes for that class. An accuracy of 0\% is assigned to any samples where no output spikes are generated.

\subsection{Recognition Accuracy}
\label{sec:sub:Results}

\begin{table}
  \textbf{\refstepcounter{table}\label{Table:Results} Table \arabic{table}.}{ Classification accuracies for N-MNIST and N-Caltech101}
\begin{center}
\begin{tabular}{|l||c|c|}
\hline
\rule{0pt}{2ex}\textbf{Task}                       &\textbf{N-MNIST}       &\textbf{N-Caltech101}    \\ \hhline{|=|=|=|}
\rule{0pt}{2ex}\textbf{Statistics kNN (k=10)}      &                       &               \\ \hline
\rule{0pt}{2ex}~~Number of events                  & 26.50\%               & 1.87\%        \\ \hline
\rule{0pt}{2ex}~~Number of ON events               & 26.11\%               & 2.06\%        \\ \hline
\rule{0pt}{2ex}~~Number of OFF events              & 26.18\%               & 1.55\%        \\ \hline
\rule{0pt}{2ex}~~Ratio of ON to OFF events         & 22.41\%               & 1.48\%        \\ \hline
\rule{0pt}{2ex}~~Mean X address of events          & 13.02\%               & 2.65\%        \\ \hline
\rule{0pt}{2ex}~~Mean Y address of events          & 13.94\%               & 2.90\%        \\ \hline
\rule{0pt}{2ex}~~Standard deviation in X address   & 24.14\%               & 3.16\%        \\ \hline
\rule{0pt}{2ex}~~Standard deviation in Y address   & 29.88\%               & 3.42\%        \\ \hline
\rule{0pt}{2ex}~~Maximum X address                 & 10.00\%               & 2.52\%        \\ \hline
\rule{0pt}{2ex}~~Maximum Y address                 & 10.00\%               & 4.32\%        \\ \hhline{|=|=|=|}
\rule{0pt}{2ex}\textbf{HFIRST}                     &                       &               \\ \hline
\rule{0pt}{2ex}~~Hard Classifier                   & 71.15\%               & -             \\ \hline
\rule{0pt}{2ex}~~Soft Classifier                   & 58.40\%               & -             \\ \hhline{|=|=|=|}
\rule{0pt}{2ex}\textbf{SKIM}                       & 83.44\%               & 8.30\%        \\ \hhline{|=|=|=|}
\rule{0pt}{2ex}\textbf{Chance}                     & 10.00\%               & 0.99\%        \\ \hline
\end{tabular}
\end{center}
\end{table}

Classification accuracies obtained by applying the methods described above to N-MNIST and N-Caltech101 are shown in Table~\ref{Table:Results}. The accuracy for each class is equally weighted when calculating the overall multiclass accuracy.

For N-MNIST, the overall number of events in each recording gives a better accuracy than looking at the number of ON or OFF events, or the ratio between them. Examples in the MNIST dataset are centered, so classification using the mean x-address and y-address only provides slightly higher accuracy than chance. Standard deviation of the x-addresses and y-addresses gives an indication of how spread out edges are in the image, with the y-address standard deviation giving the highest recognition accuracy for the kNN approaches. All MNIST examples are the same size, so classification by the maximum x-addresses and y-addresses is at chance.

For N-Caltech101, kNN classification using standard deviation of event x-addresses and y-addresses again outperforms classification using the mean address or numbers of events. However, classification using size of the example provides the highest recognition accuracy of the kNN approaches. This technique is not specific to N-Caltech101, the size of N-Caltech101 recordings depends directly on the original Caltech101 dataset, and therefore similar recognition accuracy would be achieved by looking at the size of the original frame-based images.

HFIRST performs at an accuracy of 71.15\%, which is significantly lower than the 36 class character recognition accuracy of 84.9\% reported in the original paper. However, this drop in accuracy is expected because there is far greater variation of character appearance in the N-MNIST dataset, and the HFIRST model has not been tuned or optimized for the N-MNIST dataset. HFIRST is designed to detect small objects, so it was not applied to the larger N-Caltech101 dataset.

SKIM performs best on both datasets, achieving 83.44\% on N-MNIST with 2k hidden layer neurons and 10k training iterations, and achieving 8.30\% on N-Caltech101 with 5k hidden neurons and $15\times101~=~1515$ training iterations (using 15 samples from each of the 101 categories).

\section{Discussion}
\label{sec:Discussion}

We have presented an automated process for converting existing static image datasets into Neuromorphic Vision datasets. Our conversion process uses actual recordings from a Neuromorphic sensor to ensure closer approximation of the noise and imperfections which can be expected in real-world recordings. Our conversion process also makes use of camera motion rather than motion of an image on a monitor which introduces recording artifacts (Fig.~\ref{fig:FourierBernabe}). The use of sensor motion rather than object motion is more biologically realistic, and more relevant to real world applications where most objects in the environment are stationary.

Even when objects are in motion, the velocity of these objects is typically outside of the observer's control. Sufficiently quick sensor rotations can be used to ensure that the visual motion due to sensor rotation \eqref{eq:VisualMotion} is much larger than visual motion due to the object motion. Such a scheme can be used to minimize the effect of the object motion on visual motion, and therefore on the observed intensity changes \eqref{eq:FlowConstraint}, thereby achieving a view of the object which is more invariant to object velocity.

Our conversion process allows us to leverage large existing annotated datasets from Computer Vision, which removes the need for us to gather and annotate our own data to create a dataset. Our conversion process allows Neuromorphic researchers to use data which are familiar to their Computer Vision research counterparts. We have used the conversion process described in Section~\ref{sec:conversion_process} to convert two well known Computer Vision datasets (MNIST and Caltech101) into Neuromorphic Vision datasets and have made them publicly available online.

To our knowledge, the N-MNIST and N-Caltech101 datasets we have presented in this paper are the largest publicly available annotated Neuromorphic Vision datasets to date, and are also the closest Neuromorphic Vision datasets to the original frame-based MNIST and Caltech101 datasets from which they are derived. Our conversion process allows us to easily convert other large frame-based datasets, but the time required for conversion scales linearly with the number of samples in the dataset. A 1M image dataset would take almost 6 days to convert, which is still reasonable considering that the system can be left to operate unattended. However the conversion process can become impractical for ultra-large datasets such as the 100M image Yahoo Flickr Creative Commons dataset \citep{YahooFlickr} which would take almost 1.6 years to convert.

As a starting point in tackling the datasets presented in this paper, we have provided recognition accuracies of kNN classifiers using simple statistics of the recordings as features (Section~\ref{sec:sub:kNN}), as well as accuracies using the SKIM (Section~\ref{sec:sub:SKIM}) and HFIRST (Section~\ref{sec:sub:HFIRST}) algorithms. Our aim in this paper has been to describe the dataset conversion process and create new datasets, so we have not modified or optimized the original recognition algorithms. The accuracies presented in Section~\ref{sec:sub:Results} should therefore be regarded as minimum recognition accuracies upon which to improve. Importantly, the results on both of these datasets leave a plenty of room for improvement, and we hope these datasets remain of use to the biologically inspired visual sensing community for a long time to come.

For the biologically inspired visual sensing community, we view it as important to shift from the use of stationary sensors to mobile embodied sensors. Stationary organisms in nature do not possess eyes, and even if they did, these ``eyes" would not necessarily operate in the same manner as the eyes embodied in mobile organisms. Although stationary sensing applications can also benefit from the Neuromorphic approach, the largest benefit will be for mobile applications with visual sensing needs more closely matched to tasks biology has evolved to perform. We see datasets relying on sensor motion, such as the ones presented in this paper, as a necessary step towards using mobile Neuromorphic Vision sensors in real-world applications.

\bibliographystyle{frontiersinSCNS} 
\bibliography{Datasets_bibliography}
\end{document}